\documentclass{emulateapj}

\usepackage{graphicx}
\usepackage{wrapfig}
\usepackage{natbib}

\newcommand{\be}{\begin{equation}}
\newcommand{\ee}{\end{equation}}
\newcommand{\nn}{\mbox{} \nonumber \\ \mbox{} }
\newcommand{\ba}{\begin{eqnarray}}
\newcommand{\ea}{\end{eqnarray}}
\newcommand{\om}{\omega}
\newcommand{\Alfven}{Alfv\'{e}n }

\newcommand{\Bf}{{magnetic field}}
\newcommand{\Bfs}{{magnetic fields}}

\newcommand{\ms}{magnetosphere}
\newcommand{\mss}{magnetospheres}

\newcommand{\LC}{light cylinder}
\newcommand{\Lf}{Lorentz factor}

\begin{document}

\title{Faraday rotation  in fast radio bursts}
\author{Maxim Lyutikov}
\affil{
 Department of Physics and Astronomy, Purdue University, 
 525 Northwestern Avenue,
West Lafayette, IN, USA}

\begin{abstract}
Fast Radio Bursts (FBRs) show highly different polarization properties: high/small RMs, high/small circular/linear fractions. We outline a complicated picture of polarization propagation in the inner parts of the magnetars' winds, at scales $\sim$  few to hundreds of light cylinder radii. The key point is the Faraday rotation of linear polarization in highly magnetized symmetric pair plasma, a $\propto  B^2$ effect. Position angle (PA) rotation rate is maximal for propagation across the magnetic field and disappears only for parallel propagation. In the highly magnetized regime, $\om \ll \om_B$, it becomes independent of the magnetic field. Very specific properties of PA($\lambda$) (scaling of the rotation angle with the observed wavelength $\lambda$) can help identify/sort out the propagation effects. Two basic regimes in pair plasma predict PA  $\propto  \lambda$  and  $\propto \lambda^3$ (depending on the magnetic dominance); both are different from the conventional plasma's PA = RM  $ \lambda^2$. This is the main prediction of the model. A number of effects, all sensitive to the underlying parameters, contribute to the observed complicated polarization  patterns: 
streaming of plasma along magnetic field lines  near the \LC, Faraday depolarization,  effects of limiting polarization,  the associated effect of linear-circular conversion,   and synchrotron absorption.
\end{abstract}

\maketitle

\section{ POLARIZATION OF FRBs: NO CLEAR STORY}
\label{intro}
Polarization properties of FRBs defy simple classification \cite{2019MNRAS.487.1191C,2019A&ARv..27....4P}. To quote \cite{2019A&ARv..27....4P}: ``some FRBs appear to be completely unpolarized [], some show only circular polarization [], some show only linear polarization [], and some show both''.  Understanding polarization behavior is the key to understanding FRBs.

Even in the sub-set of linearly polarized FRBs, there is no clear trend:
\begin{itemize} 
\item FRB 150807 \citep{2016Sci...354.1249R}  was nearly 80\% linearly polarized, but very small RM=12 at DM=266 (in usual astronomical unites); the average inferred magnetic field  $< B >= 5 \times  10^{-8}$  G.
\item FRB 110523 \citep{2015Natur.528..523M},  RM=186,  DM=623, $<B>=5\times10^{-7}$ G
\item FRB180301 \citep{2019MNRAS.486.3636P},  RM =  $3\times 10^3$, DM=522, $<B>=6\times10^{-6}$ G.
\item FRB 121102 \citep{2018Natur.553..182M}  was 100\% linearly polarized, (varying!) RM = $10^5$, DM=559, $< B >= 2 \times 10^{-4}$  G. At the observed frequency of $\sim$  4.5 GHz this corresponds to the PA rotation by 360 radians; this is a model-independent quantity to  be explained (in a sense that a value of RM assumes a particular frequency scaling of the rotation of polarization).
\end{itemize}

We make the following conclusion: polarization model should explain not each particular observation (e.g. large RM), but should account for large variations in polarization properties, both between different sources, temporal variations in a given source, and usual polarization behavior like in FRB180301.

Accounting for temporal variation of RM, like in FRB 121102 \citep{2018Natur.553..182M} are especially demanding, as this implies that the RM comes from a relatively compact region.

In what follows we consider Faraday effects in the near wind zone, somewhat outside the \LC. Previously a number of works considered Faraday effect inside the pulsar \ms\  \cite{1979ApJ...229..348C,1986ApJ...303..280B,2000A&A...355.1168P,2012MNRAS.425..814B}. There, the Faraday effect  is highly suppressed by a  combination number of effects:  (i) relativistic motion of plasma reduces the effective plasma frame density, and stretching  of the corresponding time scale in the lab frame; (ii) since for parallel propagation in symmetric pair plasma the  Faraday effect is absent, see Eq. (\ref{5}), the contribution to the  Faraday effects comes either  from a  slight charge-disbalance or from oblique propagation \citep{1991MNRAS.253..377K}  - both producing weak contribution (due to small ``active'' density and/or small angle of propagation).

In contrast to the \mss, in the near wind zone it  is  the {\it  total plasma density} that  contributes to the Faraday effects. In addition, the Lorentz factor of the plasma is not the  large one due to the parallel motion,  $\gamma_0 \sim  10^3-10^4$, but a mild one due to the bulk acceleration (approximately  $\sqrt{\gamma_0}$,   Eq. (\ref{101})).

\section{FARADAY EFFECT IN PAIR PLASMA }

\subsection{Cold Homogeneous plasma}

Faraday rotation of linear polarization in pair plasma appears due to difference in phase velocity of two linearly polarized waves. Generally the rate of rotation of the polarization angle \citep[][Eq. (4.6)]{1965ARA&A...3..297G}
 is
\be
 \frac{d\chi}{dz}=\frac{1}{2} \frac{\om}{c} (\Delta n) 
\label{01} \ee
where $ (\Delta n) $ is the difference in the refractive index of two linear normal modes.

\begin{figure}[h]
\includegraphics[width=0.95\linewidth]{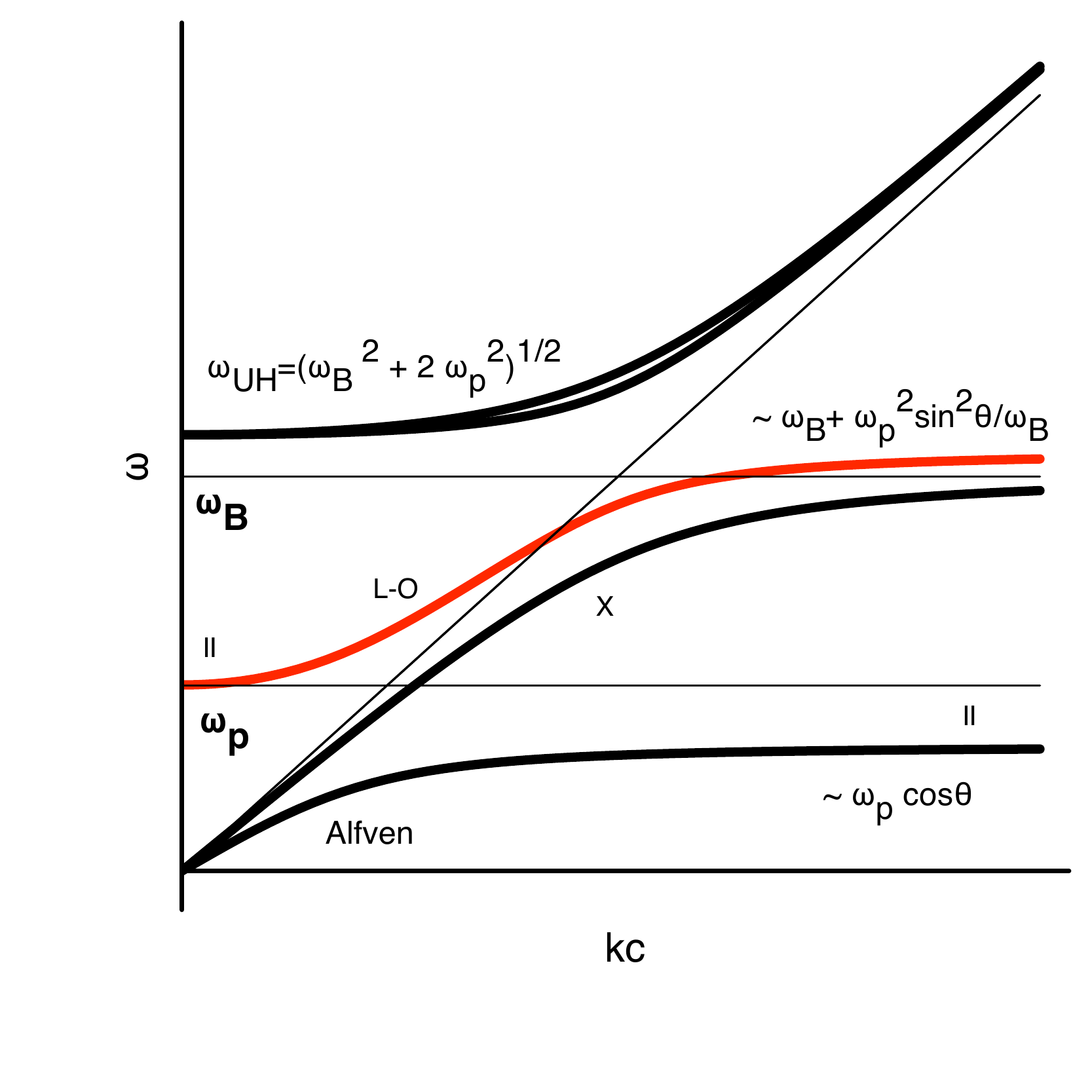}
\caption{Wave dispersions  $ \om(k) $ in pair plasma in strong \Bf, $\om_B \gg \om_p$, for oblique propagation. At low frequencies $\om \ll \om_B$ there are three modes labeled X (polarized orthogonally to   {\bf k} -{\bf B} plane), \Alfven and  L-O  (both polarized in the   {\bf k} -{\bf B} plane). The  L-O mode has a resonance at $\sim \om_B + \om_p^2 \sin^2 \theta/\om_B$ and cut-off at $\sqrt{2} \om_p$. The \Alfven mode has a resonance at  $\sim \sqrt{2} \om_p \cos \theta$. The sign $\parallel$ indicates locations where corresponding  waves are nearly longitudinally polarized. The two high frequency, $\om > \om_B$, waves with nearly identical dispersion have a cut-off at the upper hybrid frequency $\om_{UH}=\sqrt{\om_B^2 +2 \om_p^2}$,  \protect\citep{2007MNRAS.381.1190L} }
\label{1}
\end {figure}

Waves in pair plasma has been considered in a number of publications  \citep{AronsBarnard86,Kaz91,1999JPlPh..62...65L}  we follow \cite{2007MNRAS.381.1190L}. Let us consider the simplest case of cold plasma, in plasma frame. 
For  $e^\pm$ plasma in \Bf\ the  dispersion relation factorizes giving two modes: the
X mode  with the
electric vector perpendicular to the {\bf k-B} plane 
and two branches of the
longitudinal-transverse mode, which we will call 
 L-O
and Alfv\'{e}n waves,  with
 the electric vector in the {\bf k-B} plane \citep[][see Fig. \ref{1}]{AronsBarnard86}.
 X waves is a
subluminal  transverse electromagnetic wave with a dispersion relation
 \be
n^2 = 1  - \frac{ 2 \om_p^2}{  \om^2 -\om_B^2}
\label{3} \ee
here $n= kc/\om$ is refractive index, $\om_B =e B/mc $ is cyclotron frequency,  $\om_p =\sqrt{ 4\pi n_{\pm} e^2/m}$ is a plasma frequency of each species (so that for pair plasma the total plasma frequency is $\sqrt{2} \om_p$).
The \Alfven-L-O mode satisfies the 
 dispersion relation
\be
n^2 = \frac{   (\om^2 -  2 \om_p^2)(\om^2 -  2 \om_p^2 - \om_B^2) 
}{ (\om^2 -  2 \om_p^2) (\om^2 - \om_B^2 ) - 2 \om_B^2 \om_p^2 \sin^2 \theta }
\label{2} \ee
 Alfv\'{e}n branch is always subluminal  while L-O mode
is {\it superluminal} at small wave vectors and 
{\it subluminal} at large wave vectors.

In the limit $\om_p\ll \om$ we find
\ba &&
(\Delta n) = n_X-n_{A-O} = 2 \frac{\om_B ^2 \om_p^2}{\om^2 (\om^2 -\om_B^2)} \sin^2\theta = 
\nn && 
\left\{
\begin{array}{cc}
-2 \frac{ \om_B ^2 \om_p^2}{\om^4} \sin^2\theta, & \om_B \ll \om
\\
2 \frac{  \om_p^2}{\om^2} \sin^2\theta, & \om_B \gg \om,
\end{array} 
\right.
\nn &&
\frac{d \chi}{d z} =  \frac{\om_B ^2 \om_p^2}{ c\om (\om^2 -\om_B^2)} \sin^2\theta
\label{4} \ea

In homogeneous pair plasma the polarization angle at frequency $\om$ is rotated by
\ba &&
\Delta \chi = \frac{1}{2} \frac{\om}{c} \int (\Delta n)  dz=
\frac{4\pi e^3 {\rm DM} }{m_e c \om}  \sin^2 \theta \times
\nn &&
\left\{
\begin{array}{cc}
-\frac{\om_B^2}{\om^2},  & \om_B \ll \om
\\
1, & \om_B \gg \om,
\end{array} 
\right.
\nn && 
{\rm DM}= nL
\label{5} \ea
Thus, in this case $\Delta \chi  \propto {\rm DM } \times  \lambda ^1 ;  \lambda ^3$, in contrast to the conventional $ \lambda ^2$.

Unlike the case of electron-proton plasma, the Faraday effect in pair plasma disappears for parallel propagation - but only for  that special direction of propagation. Generally there is Faraday rotation $\propto B^2$. In the case of infinitely strong magnetic field there is Faraday effect, but it is independent of the value of the  magnetic field.

\subsection{Faraday effect in the inner parts of the wind}

Next we apply the above relations to the inner parts
of magnetars' (and pulsars') winds. As a starting point,
let us approximate the wind as a sequences of toroidal
magnetic loops accelerating away from the light cylinder  \citep{1969ApJ...158..727M,1970ApJ...160..971G,1973ApJ...180L.133M}, 
see Section 2 in \cite{2021arXiv210902524L} for a concise summary. This approximation neglects motion of the ejected plasma along the magnetic field lines (this is important within pulsar magnetospheres \cite{2006MNRAS.366.1539P,2000A&A...355.1168P}. Outside of the light cylinder such effects are also important, \S \ref{r0},  the model described below in this section  is applicable not for $ r \geq R_{LC}$  but from  $r \geq  r _{0}$, Eq. (\ref{101}).

Under the assumption, the EM waves propagate across magnetic field, $\theta= \pi/2$, through relativistically accelerating wind with
\be
\Gamma_w = \frac{r}{R_{LC}}
\label{21} \ee
In the frame of the wind, denoted with prime, the rate of PA rotation is
\ba && 
\left( \frac{d \chi}{d z} \right) ^\prime =   \frac{\om_B ^{\prime,2} \om_p ^{\prime,2} }{ c\om' (\om ^{\prime,2}  -\om_B ^{\prime,2} )} 
\nn &&
B'= B/\Gamma_w
\nn &&
n'= n/\Gamma_w
\nn &&
\om '= \om/\Gamma_w
\label{6} \ea
In the lab frame
\be
 \frac{d \chi}{d z} = \left( \frac{d \chi}{d z} \right) ^\prime /\Gamma_w
\label{22} \ee
Let us parametrize the properties of the wind by wind luminosity $L_w$ and the ratio of Poynting to particle fluxes:
\ba &&
\mu_w =\frac{L_w}{ \dot{N}  m_e c^2} =  \frac{B_{LC}^2}{4\pi n_{LC} m_e c^2} 
\nn &&
B_{LC}= \frac{\sqrt{L_w} \Omega}{c^{3/2}}
\nn &&
 n_{LC} = \frac{L_w}{4\pi \mu m_e c^3 R_{LC}^2 }
\label{7} \ea
(terminal Lorentz factor of the wind is $\gamma_w = \mu^{1/3}$).

Thus, in the acceleration zone
\ba &&
n' =\left( \frac{R_{LC}}{r} \right)^{-3} n_{LC} 
\nn && 
B' = \left( \frac{R_{LC}}{r} \right)^{-2} B_{LC}
\nn &&
\sigma(r) = \frac{B^{\prime,2}}{4 \pi n' m_e c^2} =  \frac{R_{LC}}{r} \mu_w
\label{sigma1}
\label{8} \ea
(subscript LC indicates quantities measured at the light cylinder).

For a given frequency $\om$ the cyclotron resonance occurs at
\ba && 
r_B = \frac{e \sqrt{L_w}}{c^{3/2} m_e \om}
\nn && 
\frac{r_B }{R_{LC}} = 30 L_{w, 38} \nu_9^{-1} P^{-1}
\label{9} \ea
where period is in seconds.

The polarization rotation rate, as measured in lab frame is then
\ba && 
\frac{d \chi}{dr} = 
\left\{
\begin{array}{cc}
 \frac{\om_p^{\prime,2} } {c \om}  &r  \leq  r_B
\\
- \Gamma_w^2 \frac{\om_B ^{\prime,2} \om_p ^{\prime,2} }{ c\om {\prime,3} }  &r  \geq  r_B
\end{array}
\right.
 =
\nn &&
\frac{e^2 L_w R_{LC}} { c^4 \mu r^3 \om m_e^2} \times
\left\{
\begin{array}{cc}
1 &r  \leq  r_B
\\
\frac{ e^2 L_w}{ c^3 r^2 \om^2 m_e^2}  &r  \geq  r_B
\end{array}
\right.
\label{10} \ea
Thus, we have two regimes for the PA rotation,  $\propto \lambda$ at $ r\leq r_B$  and $\propto \lambda^3$ , at $ r\geq r_B$.
The region near the cyclotron resonance $ r_B$ presents a challenge, both in terms of the possibility of cyclotron absorption, and due to large rates of PA rotation. If cyclotron absorption is negligible, \S \ref{Cyclotron} , large rotation angle near the resonance will be mostly cancelled, since at two sides of the resonance the rotation direction is in the opposite sense. (But small mismatch between inner and outer parts may produce large net rotation near the resonance.)
In both regions, $ r\leq r_B$  and just outside the $r_B$, the rotation is fastest near the inner boundary. Next we consider the most interesting case, when PA rotation is acquired mostly before the cyclotron resonance

\subsection {Plasma motion along magnetic field near light cylinder} 
\label{r0}

Particles leaving the pulsar/magnetar magnetosphere have large parallel momentum. Let them move near the light cylinder with the parallel Lorentz factor $\gamma_{0}$. Conservation of angular momentum then determines the evolution of (properly defined) parallel Lorentz factor $\gamma_\parallel$:
\ba &&
\gamma_{0} R _{LC}= \gamma_{tot} r
\nn &&
\gamma_{tot}= \gamma_\parallel \Gamma_w
\label{11} \ea

Hence  $\gamma_\parallel$: becomes of the order of unity at
\ba &&
r_{0} = \sqrt{\gamma_{0}} R_{LC}
\nn &&
\Gamma_w (r_{0}) = \sqrt{ \gamma_{0}}
\label{101} \ea
This is the smallest radius where the model becomes applicable (for smaller  $r$ large relativistic plasma velocity along magnetic field, in combination with the small angle of the waves with respect to the magnetic field suppresses Faraday rotation).

We estimate the radial integral of the first term in (\ref{10})
\be
\Delta \chi = \frac{ e^2 L_w}{2 \gamma_0 \mu  m_e^2  c^4 R_{LC} \om} =
200 L_{w,38} P^{-1} \gamma_{0,2} ^{-1} \mu_6 ^{-1} \lambda \, {\rm rad}
\label{12} \ee
This is our one of the main results: Faraday polarization rotation in the wind can be large. 
Condition $\Delta \chi \ggg 1$  leads to Faraday depolarization.

This values is close to the maximal observed one, \S \ref{intro}
\citep{2018Natur.553..182M}.
Rotation angle can be large, but  is highly dependent
on the assumed parameters, especially $\mu$ and $\gamma_{0}$. Rotation angle in this region is  $\propto \lambda$.
Also, the cyclotron resonance condition (\ref{9}), which assumes Michel's solution, requires $r_B \geq r_{0}$,
\be
\gamma_{0} \leq \gamma_{0,crit}= \frac{ e^2 L_w}{   m_e^2  c^3  \om^2 R_{LC}^2}
\label{13} \ee
It might realistically  be violated. If this condition is violated, the cyclotron resonance will be moved out (due to extra reduction of the wave's frequency in the plasma frame). 
The polarization angle  (\ref{12}) will be reduced.

\section{COMPLICATIONS: LIMITING POLARIZATION, PRODUCTION OF CIRCULAR COMPONENTS AND
CYCLOTRON ABSORPTION}

\subsection{ Limiting polarization radius}
\label{limiting}
Separation of modes into X and L-O branches may be
violated if the rate of change of plasma parameters is sufficiently fast, so that the mode propagation becomes non-adiabatic \cite[the effect of limiting polarization][]{1952RSPSA.215..215B}. This occurs when the wavelength of the beat between two modes becomes larger than the scale at which the properties of the modes change.  In our case his condition becomes
\be
\left( \frac{\om}{\Gamma_w c} \right) \left( \frac{r}{\Gamma_w } \right) \left( \Delta n\right)^\prime \geq 1
\label{23} \ee
This is a condition that propagation is adiabatic: it is satisfied for radii less than the limiting polarization
radius $R_{LP}$:
\be
\frac{ r}{R_{LC}} \leq\frac{  R_{LP}}{R_{LC}}  = \frac{\sqrt{2} e \sqrt{L_w}  }{m_e c^2  \sqrt{\mu} \sqrt{\om}  \sqrt{R_{LC}} }= 280  L_{w,38}^{1/2}  P^{-1/2}  \mu_6 ^{-1/2} \lambda^{1/2}
\label{14} \ee

Ratio of the cyclotron absorption radius $r_B$ (\ref{9})  to limiting polarization $R_{LP}$ (\ref{14})  
\be
\frac{r_B} {R_{LP}} \approx \sqrt{ \frac{\mu \Omega} {\om_B}}
\ee
It is smaller than unity for 
\be 
\mu \leq \frac{\om}{\Omega} \to \gamma _w \equiv \mu^{1/3} \leq \left( \frac{\om}{\Omega}\right)^{1/3} 
\ee

\begin{figure}[h!]
\includegraphics[width=0.95\linewidth]{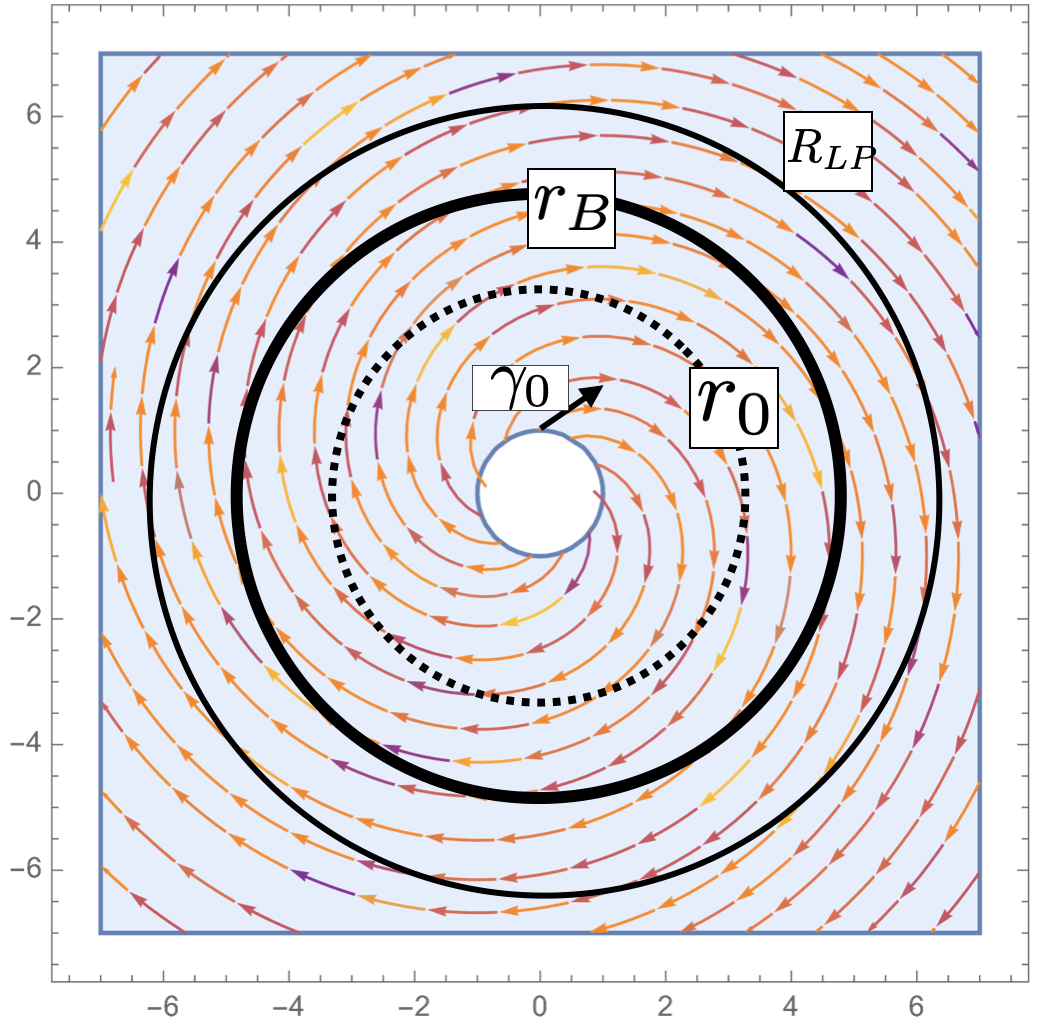}
\caption{ Geometry of Faraday rotation in the inner parts of the wind (not to scale). Arrows are \Bf\  \protect\citep{1973ApJ...180L.133M}. Particles leave the \ms\ with \Lf\ $\gamma_0$, moving along the \Bf\ lines. At distance $r_0 \sim \sqrt{\gamma_0} R_{LC}$ the parallel momentum becomes non-relativistic. At radii smaller than $r_0$ wave-plasma interaction effects are suppressed by large parallel momentum of particles; at radii larder than $r_0$ the wind can be described as a radially accelerating set of toroidal \Bfs\ carrying cold plasma. Cyclotron resonance occurs at $r_B$, limiting polarization radius is  $R_{LP}$. Depending on parameters of the flow relative location of $r_0$, $r_B$ and $R_{LP}$ may change.}
\label{Faraday-pair-pic}
\end {figure}

\subsection{Production of circular polarization}
When a wave reaches the limiting polarization radius, circular polarization will be produced \citep{2000MNRAS.311..555H,2012MNRAS.425..814B,2020MNRAS.498.5003J}. 
A total rotation angle on the Poincare  sphere, and hence the amount of the produced circular component,  can estimated as  angle that the magnetic field spiral makes with the $\phi$  direction at the location of the limiting polarization radius  $R_{LP}$. The expected Stokes' $V$ is then
\be
V\sim \frac{R_{LC}}{R_{LP}}\sim  \, {\rm few} \% 
\label{V} \ee

\subsection{Cyclotron absorption in the wind}
\label{Cyclotron} 
The resonant  optical depth can be  estimated as \citep[][]{1996ASSL..204.....Z,1994ApJ...422..304T,lg06}
\ba && 
\tau_{res} \approx \sigma_{res}   n'\frac{r}{\Gamma}
\nn &&
\sigma_{res} = \frac{\pi^2 e^2  }{m_e c \om_B'}
\nn &&
\tau_{res} =\frac{ e \sqrt{L_w}}{ m_e c^{3/2} \mu r \Omega} \approx
\frac{1}{\mu} \frac{\om}{\Omega} \times
\left\{
\begin{array}{cc}
1 &r  \leq  r_B
\\
\left(\frac{\gamma_{0} }{\gamma_{LC,min} }  \right)^{-1/2}   &r  \geq  r_B
\end{array}
\right.
\label{15} \ea
Thus, for sufficiently high initial parallel Lorentz factor, $\gamma_{0}  \gg \gamma_{LC,min} $ and large $\mu \gg 1 $, cyclotron absorption can be avoided.

\section{DISCUSSION}

In this letter we discuss the properties of polarization transfer in the near wind regions of magnetars, presumed {\it loci} of FRBs; magnetospheric model of radio emission from magnetars and FRBs \citep{2002ApJ...580L..65L,2013arXiv1307.4924P,2020arXiv200505093L,LyutikovFEL} is assumed.  \citep[In particular,  the model of][allows for the intrinsic   production of both circular and linearly polarized waves, with clear correlation between spectral and polarization properties - multiple spectral  stripes come with linear polarization]{LyutikovFEL}.s

We concentrate on the inner parts of the wind,  from the light cylinder to $\sim$  hundreds of the light cylinder distances. We find a  complicated picture, that cannot be expressed as a combination of just a few parameters: the neutron stars surface magnetic field and period, multiplicity of plasma production in the magnetosphere, the resulting bulk Lorentz factor along the fields lines near the light cylinder, and the structure of the inner acceleration region of the wind all contribute sensitively to the evolution of the polarization. The key effect, which seems to be often misinterpreted, is the Faraday rotation in pair plasma, a $B^2$ effect. In most space/laboratory applications, in the limit $\om_B,\om_p\ll \om$ it is often ignored. In the highly magnetized pair plasma it does produce rotation of polarization, even independent of the magnetic field in the high-magnetic field limit.

 We outline a complicated multi-parameter problem:  (i) Faraday rotation  can be large, Eq, (\ref{12});  (ii) very large   Faraday rotation can lead to depolarization; (iiI) large parameter $\mu$ (related to the final \Lf\ of the wind, $\gamma_w = \mu^{1/3}$ leads to small Faraday rotation; (iv) large initial \Lf\ decreases Faraday rotation, Eqns. (\ref{12})-(\ref{13}); (v) cyclotron resonance (absorption and fast rates of Faraday rotation) may be important, \S \ref{Cyclotron}. (vi) effects of limiting polarization, \S \ref{limiting}, may lead to the production of circular component at $\sim$ few $ \%$ level.  We suggest that this multi-parameter problems leads to highly heterogeneous observations of FRB's polarization. 

The main prediction of the model is that scaling of the PA with frequency should deviate from the conventional $ \chi \propto \lambda^2$. Our  ``bare-bone''  model predicts  $\chi \propto \lambda$.
There are observational hints, the most interesting analysis is by \cite{2019MNRAS.486.3636P}, their Fig. 8. They demonstrate that {\it linear} relation $\chi \propto \lambda$ is consistent with data.

The astrophysical (pulsars's, FRB's and probably AGN and GRB jets) consequences cannot be simply quantified: different scalings of the rotations rates with wavelength for $\om \leq \om_B$ and  $\om\geq \om_B$, effects of limiting polarization, cyclotron absorption, and very fast PA rotation rates near the cyclotron resonance, make the model sensitive to the particular parameters of the source.
We take this sensitivity to the parameters as a success of the model: the observed polarization patterns in FRBs are  highly variable: the present model accounts both the variability, and maximal values of the PA swing, Eq. (15) comparable to the ones observed by \cite{2018Natur.553..182M}.
Also, origin of polarization properties in the
near wind naturally explains temporal variations of polarization properties due to the wind non-stationarity.

On the other hand, pulsars clearly do not show such wild polarization behavior.  The rotating vector model \citep{1969ApL.....3..225R}, that neglects all the propagation effects, does account for many pulsar PA profiles (though there are many exceptions when it's not: e.g. in Crab).  Relation (\ref{12}) gives the simplest estimate of the Faraday effect in the wind. The propagation effects are easily ``killed'' by the low plasma density expected in the wind: smaller density  (larger $\mu$ parameter)  in pulsars  than in magnetars  (the \cite{GJ} density   is much smaller that what is expected in magnetars \citep{tlk,2007ApJ...657..967B}). 

\section{ACKNOWLEDGEMENTS}

This work had been supported by NASA grants 80NSSC17K0757 and 80NSSC20K0910, NSF grants 1903332 and 1908590.
I would like to thank Vasily Beskin  and Kiyoshi Masui for discussions.

\section{DATA AVAILABILITY}
The data underlying this article will be shared on reasonable request to the corresponding author.

 \bibliographystyle{apj} 
 \bibliography{/Users/maxim/Home/Research/BibTex}

\end{document}